# An empirical study on the impact of IDE tool support in pair and solo programming

Omar S. Gómez
Facultad de Informática y Electrónica
Escuela Superior Politécnica de Chimborazo
Riobamba, Ecuador
ogomez@espoch.edu.ec

***Abstract***: Agile software development has been widespread adopted. One well-known agile approach is eXtreme Programming (XP) where pair programming (PP) is a relevant practice. Although various aspects of PP have been studied, we have not found, under a traditional model of PP, studies that examine the impact of using an IDE tool support. In an attempt to obtain a better understanding of the impact of using an IDE, we present the results of a controlled experiment that expose the influence on quality, measured as the number of defects injected per hour, and cost, measured as the time necessary to complete programming assignments, of pair and solo programming with and without the use of an IDE. For quality, our findings suggest that the use of an IDE results in significantly higher defect injection rates (for both pairs and solos) when the programming assignment is not very complicated. Nevertheless, defect injection rates seem to decrease when pairs work on more complicated programming assignments irrespective of the tool support used. For cost, the programming assignment significantly affects the time necessary to complete the assignment. Finally, both aspects (quality and cost) are affected in a similar manner when either pair or solo programming is used.

## I. INTRODUCTION

Currently, the adoption of agile approaches in the development and maintenance of software products is becoming relevant in the software industry. The successful adoption of agile approaches has been reported in several countries [1-7].

One of the first and well-known agile approaches is eXtreme Programming, or XP [8,9], which focuses on twelve practices for software development: planning games, short releases, system metaphors, simple design, testing, refactoring, pair programming, collective ownership, continuous integration, 40-hour-week, on-site customers, and coding standards.

Pair programming is a common practice used either independently or as part of XP, in this practice, two programmers work together on the same task using a computer. One of the programmers (the driver) writes the code, and the other (the observer) actively reviews the work performed by the driver. Essentially, the observer reviews the work for possible defects, writes down notes, or defines strategies for solving any issue that can arise in the task that they are working on.

Various empirical studies that report beneficial effects of the use of this practice have been conducted [10-24]. Some beneficial effects reported in these studies are that pair programming helps to produce shorter programs and achieve better designs; programs contain fewer defects than those written individually, and pairs usually require less time to complete a task than do programmers working individually.

Although different aspects of pair programming have been studied, we have not found works that examine the impact of using a tool support such as an integrated development environment (IDE). With the objective of obtaining a better understanding of the influence of the use of an IDE with this practice, in this work, we present the results of an empirical study in the form of a controlled experiment that investigates the effect on quality and cost of pair and solo programming with and without the use of an integrated development environment (IDE).

The remainder of the document is organized as follows: In Section II, we describe the background; in Section III, we present the experimental setting; in Section IV, the analysis and results are presented; in Section V, we discuss the findings; and finally, in Section VI, the conclusions are drawn.

## II. BACKGROUND

The use of a tool support for pair programming has been studied under a distributed model, wherein pairs are located at different geographical locations and are working in a synchronous or asynchronous manner [25-31]. We have not found studies from other authors that investigate the effect of using a tool support under a traditional pair programming model, wherein pairs work in the same location with the same computer simultaneously.

The experiment reported here has its origins in the work of [10]. In [10], the authors conducted a controlled experiment as part of a course on design of experiments (DoE) in software engineering (SE). The experiment was conducted at the Faculty of Mathematics of the Autonomous University of Yucatan in the Software Engineering degree program. In this experiment, a Latin square design [32] was used with the aim of blocking the programs being coded and the tool support used, thus analyzing only the type of programming (pair and solo).

The main characteristics of a Latin square design are that there are two blocking factors. The treatment is present at each level of the first blocking factor as well as at each level of the second blocking factor. This design is arranged with an equal number of rows (blocking factor one) and columns (blocking factor two). Treatments are represented by Latin symbols, where each symbol is present exactly once in each row and exactly once in each column. An example of the structure of this design is shown in Table I.

TABLE I
Example of a Latin square experimental design

| A | B | C |
|---|---|---|
| B | C | A |
| C | A | B |

In this type of experimental design, blocking is used to systematically isolate the undesired source of variation in the comparison of treatments. In the case of the experiment reported in [10], a Latin square design was used in an attempt to block the program being coded and the tool support, thus reducing the undesired source of variation between the treatments of interest, namely, the pair and solo programming approaches. The Latin square design structure used in [10] is shown in Table II.

TABLE II
Latin square experimental design used in [10]

| Program / tool support | IDE | Text editor |
|---|---|---|
| Calculator | Solo programming | Pair programming |
| Encoder | Pair programming | Solo programming |

The aspects studied in [10] were duration (cost) and effort of 7 pairs and 7 solos who coded two programs with and without the use of an IDE. The duration was measured as the time in minutes necessary to code the programming assignment, whereas effort was measured as the amount of labor spent on coding the programming assignment (measured in person-minutes).

With respect to duration, the authors in [10] reported a significant (at $\alpha$=0.1) decrease in time of 28% in favor of pairs and a medium effect size $d$=0.65.

Regarding effort, the authors in [10] reported a significant (at $\alpha$=0.1) decrease in effort of 30% in favor of solos and a medium effect size $d$=0.64.

The authors observed a difference with respect to the time spent for coding both programs. Because the program and tool support were treated as blocking factors, it is important to note that with the used Latin square design, it was not possible to assess possible interactions between the treatment (type of programming) and the blocking factors (program being coded and tool support). The variability observed in both programs suggests the need for further investigation of a possible influence between these two variables that were used as blocks and the treatment of interest.

## III. EXPERIMENTAL SETTING

Considering the issues discussed in the previous section, we decided to conduct another controlled experiment, therein varying the experimental design as well as the effect operationalizations. According to [33,34], the experiment reported here can be considered as an operational replication. The experimental design is varied with the objective of examining a possible influence based on how either the tool support used for coding or the program used to code (programming assignment) may affect the type of programming (pair or solo).

With respect to the effect operationalization, one of the aspects studied in [10] was the time (cost) that pairs and solos spent coding the programming assignments. We keep this aspect; in addition, we examine another aspect related to the quality of the software products (programming assignments) produced by the pairs and solos.

Following the Goal-Question-Metric approach [35], which facilitates the identification of the object of study, purpose, quality focus, perspective and context of an experiment, we define the experiment reported here as follows:

"Study pair and solo programming with the purpose of evaluating how quality and cost could be affected by the use of an IDE or a text editor as tool support along with the program that the programmers code. This study is conducted from the point of view of the researcher within an academic context. This context is composed of junior-year students enrolled in a course on DoE, where they will code, in pairs or individually, two programming assignments with a different tool support."

Based on the previous experimental definition, we derive the following hypotheses:

- **H0$_a$**: Participants working in pairs and individually develop software products with similar quality.
- **H0$_b$**: Coding through an IDE and a simple text editor yield software products with similar quality.
- **H0$_c$**: Implementing the specification of program A and program B yield a software product with similar quality.
- **H0$_d$**: Software product quality is not affected by the relationship between the type of programming (pair or solo) and the type of tool support used.
- **H0$_e$**: Software product quality is not affected by the relationship between the type of programming and the type of program being coded.
- **H0$_f$**: Software product quality is not affected by the relationship between the type of tool support used and the type of program being coded.
- **H0$_g$**: Software product quality is not affected by the relationship between the type of programming, type of tool support and type of program being coded.
- **H0$_h$**: Participants working in pairs and individually spend similar time (cost) coding the programming assignments.
- **H0$_i$**: Coding using an IDE and a simple text editor is performed in a similar amount of time (cost).
- **H0$_j$**: The implementation of the specification of program A and program B is performed in a similar amount of time (cost).
- **H0$_k$**: Cost is not affected by the relationship between the type of programming (pair or solo) and the type of tool support used.
- **H0$_l$**: Cost is not affected by the relationship between the type of programming and the type of program being coded.
- **H0$_m$**: Cost is not affected by the relationship between the type of tool support used and the type of program being coded.
- **H0$_n$**: Cost is not affected by the relationship between the type of programming, tool support and type of program being coded.

*A.    Experimental Design*

The defined hypotheses for this experiment will be tested using different measurements collected from the participants of this experiment. The collected measurements belongs to four groups: (1) Participants working in pairs on programs A and B with an IDE, (2) participants working in pairs on programs A and B with a simple text editor, (3) participants working individually on programs A and B with an IDE, and (4) participants working individually on programs A and B with a simple text editor. With the collected measurements, we will contrast them with the defined hypotheses.

With the goal of collecting the maximum number of measurements possible, we used a factorial design with repeated measurements. A factorial design enables the study of several factors and the interactions among them. In this experiment, we applied a $2^2$ factorial design and selected the main factors as the type of programming (pair and solo) and the tool support (IDE and text editor). This design is then repeated (maintained on the same experimental unit) along with the programs being coded. As mentioned in [36], repeated measurements on the same experimental units (in this case, groups of pairs and solos) provide an efficient use of resources compared to extracting measurements from different experimental units. Another characteristic of this experimental design is the reduction of the variance of estimates, thereby allowing statistical inferences to be made with fewer participants.

*B.    Participants, Tasks and Objects*

The participants for this experiment were junior-year students (i.e., in their third year) enrolled in a DoE course in a software engineering program at Autonomous University of Yucatan. The experiment reported here was conducted during the summer semester of 2013, and it was scheduled to end in November 2013. In this experiment, there were 24 students (8 pairs and 8 solos) in total who assisted and finished all the programming assignments (out of a total of 28 students). According to the Dreyfus and Dreyfus programming expertise classification [37], we categorized participants as advanced beginners; the participants have working knowledge of key aspects of Java programming practice. On average, they reported having 1.94 (SD=0.97) years of experience with the Java programming language and 1.81 (SD=1.03) years of experience with the NetBeans IDE. Verbal consent was obtained from students, and participants were informally advised about the data retained and that anonymity was fully ensured. No sensitive data were collected for the experiment.

From a total of 356 credits, which is the minimum number of credits necessary to complete the SE degree curricula, at the time of the experiment, the participants had completed on average 213.40 (SD=49.92) credits of the SE degree (59.94% complete). Regarding the gender of the participants, there were 22 males and 6 females. Table III shows the gender distribution for each experimental unit, the identification number of each participant and the number of measurements performed. A total of 3 out of 19 experimental units did not attend all the planned experimental sessions (two sessions); therefore, only one measurement per aspect (quality and cost) was collected from these three experimental units.

TABLE III
Characteristics of the experimental units (eu)

| Experimental unit (eu) | Composition | Id participant | Measurements performed |
|---|---|---|---|
| $eu_1$ | Male, Male | 22, 5 | 2 |
| $eu_2$ | Male, Male | 27, 17 | 2 |
| $eu_3$ | Female, Female | 6, 24 | 2 |
| $eu_4$ | Male, Male | 25, 18 | 1 |
| $eu_5$ | Male, Male | 14, 3 | 2 |
| $eu_6$ | Female, Female | 28, 7 | 2 |
| $eu_7$ | Male, Male | 12, 4 | 2 |
| $eu_8$ | Male, Male | 8, 19 | 2 |
| $eu_9$ | Male, Male | 9, 11 | 2 |
| $eu_{10}$ | Male | 16 | 1 |
| $eu_{11}$ | Male | 2 | 2 |
| $eu_{12}$ | Male | 10 | 1 |
| $eu_{13}$ | Male | 15 | 2 |
| $eu_{14}$ | Female | 1 | 2 |
| $eu_{15}$ | Male | 23 | 2 |
| $eu_{16}$ | Male | 20 | 2 |
| $eu_{17}$ | Male | 26 | 2 |
| $eu_{18}$ | Female | 21 | 2 |
| $eu_{19}$ | Male | 13 | 2 |

Participants were randomized according to gender and allocated into four groups (treatment combinations) according to the $2^2$ factorial design: (1) a group of pairs working with an IDE, (2) a group of pairs working with a text editor, (3) a group of solos working with an IDE, and (4) a group of solos working with a text editor. According to the repeated measurement design, all the participants coded two programs in two different sessions. Before the experiment was conducted, we presented a talk to the students about eXtreme Programming with a special focus on pair programming. In this talk, we explained the main concepts of this programming practice and how it can be applied. In another talk, we reinforced the concepts of pair programming and explained how to compile and run a Java program using only a text editor and the operating system console. Finally, we explained to students how to collect the measurements during the experimental sessions. The collection procedure consisted of writing down the amount of time that students spent writing a program (we asked them to record starting and ending times and compute the difference in minutes). In addition, we asked them to record only logical defects that they injected during coding. We explained to students two basic types of defects that they can commit: syntax and logical defects. In a syntax defect, the program cannot be compiled; this is especially important for students working with a simple text editor because of they do not receive syntax hints from the IDE. On the other hand, when a logical defect is committed, the program can be compiled and run, but it will not work properly, i.e., it will not function according to its specification. As tool support, pairs and solos utilized either the NetBeans IDE or a simple text editor (Notepad, Pico or Nano) with the Java programming language. Printed forms were available for time and defect registration.

Prior to the experimental sessions, a training phase was conducted. The training phase enables additional control over experimental conditions, reducing undesired variations in the measurements. In two separate training sessions, participants coded two programs. This training phase allowed the pairs to be immersed and to gain experience with the pair programming practice. The experience gained for pairs alleviates the issues discussed in pair programming experiments regarding the lack of training that pairs receive in comparison to solos programmers [38].

We wrote the specification for two console programs that participants could code, compile, run and test during each experimental session. For the first program (identified as encoder, or program A), we asked participants to code a simple encoding-decoding program. Given a specified table that contains letter switches, the program must be able to encode or decode a line of text. The program receives two arguments: one that indicates whether to encode or decode the text and one that indicates the line of text (enclosed by quotation marks) to process.

For the second program (identified as calculator, or program B), we asked participants to write a simple calculator that evaluates expressions containing decimal numbers along with the operators addition, subtraction, multiplication, and division. If the expression is valid, the program prints the results on the screen; otherwise, the program prints a message indicating an invalid expression.

*C.     Experimental Conduct*

Once the random assignment of participants to treatment combinations (pairs using an IDE, pairs using a text editor, solos using an IDE, and solos using a text editor) was performed prior to the training phase, the experimental units (pairs and solos) worked with the same treatment combination (type of programming and tool support assigned) during the training and experimental sessions, varying only the program being coded in each session.

The allotted time for each experimental session was 90 minutes. Both sessions were conducted in one of the computer classrooms of the university. Once most of the students were in the classroom, we started each session. In the first experimental session, we gave participants directions and projected the specification of the program to code (program A, encoder) onto the screen. In this session, two solo programmers spent more time coding than assigned to the session, namely, 107 and 147 minutes. Because the first of these two participants was nearly finished with the programming assignment, we decided to ask them to wait. We asked the second participant to pause their programming activities and restart them at home (while properly performing the measurement collection process).

The second experimental session was conducted in a similar manner as the previous session; we gave the participants directions and projected the second specification (program B, calculator) onto the screen. In this session, half of the experimental units (pairs and solos) finished the programming assignment on time. For the other half, we scheduled an extra session in the same classroom. In this extra session, two solos and one of the pairs did not finish the programming assignment; thus, we asked them to finish the assignment at home.

The programming assignment for each session was considered as finished or completed once we verified (in each experimental unit) the proper operation of the program against its specification.

*D.     Metrics*

The metrics operationalized for the effect constructs (quality and cost) were defined as the number of defects injected per hour (product quality) and the elapsed time in minutes that participants spent coding the programming assignment until it ran according to its specifications (cost). Following the suggestions in [39], we evaluated the product quality based on an external metric. It seems that the use of internal metrics for assessing product quality can lead to unreliable results [40,39].

After the experimental sessions were finished, we collected measurements from 28 participants; however, we had to omit the measurements of three experimental units (two solos and one pair) because the participants did not attend experimental session two. For the applied experimental design, it is necessary to collect all the repeated measurements from the experimental units.

With respect to the number of defects injected per hour (product quality), the following measurements were collected:

For session 1 (program A, encoder)
- Pairs working with an IDE: 2.86 ($eu_1$), 4.29 ($eu_2$), 9.60 ($eu_3$), 4.14 ($eu_5$).
- Pairs working with a simple text editor: 0.92 ($eu_6$), 3.43 ($eu_7$), 1.33 ($eu_8$), 4.39 ($eu_9$).
- Solos working with an IDE: 12.86 ($eu_{11}$), 5.22 ($eu_{13}$), 2.24 ($eu_{14}$).
- Solos working with a simple text editor: 2.14 ($eu_{15}$), 1.25 ($eu_{16}$), 2.86 ($eu_{17}$), 0.41 ($eu_{18}$), 1.00 ($eu_{19}$).

For session 2 (program B, calculator)
- Pairs working with an IDE: 0.96 ($eu_1$), 1.30 ($eu_2$), 0.62 ($eu_3$), 1.06 ($eu_5$).
- Pairs working with a simple text editor: 0.87 ($eu_6$), 1.68 ($eu_7$), 0.85 ($eu_8$), 1.15 ($eu_9$).
- Solos working with an IDE: 6.67 ($eu_{11}$), 5.71 ($eu_{13}$), 0.63 ($eu_{14}$).
- Solos working with a simple text editor: 0.67 ($eu_{15}$), 1.24 ($eu_{16}$), 1.80 ($eu_{17}$), 0.83 ($eu_{18}$), 0.67 ($eu_{19}$).

Regarding the elapsed time in minutes required to code the programming assignments (cost), the following measurements were collected:

For session 1 (program A, encoder)
- Pairs working with an IDE: 42 ($eu_1$), 14 ($eu_2$), 25 ($eu_3$), 58 ($eu_5$).
- Pairs working with a simple text editor: 65 ($eu_6$), 70 ($eu_7$), 45 ($eu_8$), 41 ($eu_9$).
- Solos working with an IDE: 28 ($eu_{11}$), 23 ($eu_{13}$), 107 ($eu_{14}$).
- Solos working with a simple text editor: 56 ($eu_{15}$), 48 ($eu_{16}$), 63 ($eu_{17}$), 147 ($eu_{18}$), 60 ($eu_{19}$).

For session 2 (program B, calculator)
- Pairs working with an IDE: 125 (eu$_1$), 92 (eu$_2$), 97 (eu$_3$), 340 (eu$_5$).
- Pairs working with a simple text editor: 138 (eu$_6$), 143 (eu$_7$), 211 (eu$_8$), 104 (eu$_9$).
- Solos working with an IDE: 36 (eu$_{11}$), 63 (eu$_{13}$), 479 (eu$_{14}$).
- Solos working with a simple text editor: 89 (eu$_{15}$), 97 (eu$_{16}$), 100 (eu$_{17}$), 432 (eu$_{18}$), 180 (eu$_{19}$).

## IV.  ANALYSIS AND RESULTS

In this section, we present both descriptive and inferential statistics for the collected measurements. In addition, we present results from a qualitative analysis referring to a questionnaire that participants responded to on their pair programming experience. Tables IV and V show the descriptive statistics for the programming type, tool support and programming assignment with respect to product quality and cost.

TABLE IV
Descriptive statistics with regard to number of defects injected per hour (quality) for programming type, tool support and program being coded

| Factor | Level | $n$ | mean | SD | Min. | Max. |
|--------|-------|-----|------|-----|------|------|
| Programming type | Pair | 16 | 2.47 | 2.34 | 0.62 | 9.60 |
| | Solo | 16 | 2.89 | 3.30 | 0.41 | 12.86 |
| Tool support | IDE | 14 | 4.15 | 3.65 | 0.62 | 12.86 |
| | Text editor | 18 | 1.53 | 1.06 | 0.41 | 4.39 |
| Program | Encoder | 16 | 3.68 | 3.32 | 0.41 | 12.86 |
| | Calculator | 16 | 1.67 | 1.81 | 0.62 | 6.67 |

TABLE V
Descriptive statistics with respect to elapsed time (cost, in minutes) for programming type, tool support and program being coded

| Factor | Level | $n$ | mean | SD | Min. | Max. |
|--------|-------|-----|------|-----|------|------|
| Programming type | Pair | 16 | 100.62 | 81.94 | 14 | 340 |
| | Solo | 16 | 125.50 | 135.75 | 23 | 479 |
| Tool support | IDE | 14 | 109.21 | 134.60 | 14 | 479 |
| | Text editor | 18 | 116.06 | 92.73 | 41 | 432 |
| Program | Encoder | 16 | 55.75 | 33.10 | 14 | 147 |
| | Calculator | 16 | 170.38 | 131.79 | 36 | 479 |

Regarding product quality (Table III), on average, both types of programming appear to produce similar defect injection rates. In the case of tool support, participants who worked with an IDE seem to have higher defect injection rates. With respect to the program being coded, it seems that participants had higher defect injection rates when coding the encoder program.

In the case of cost (Table IV), solo programmers seem to spend more time coding than do pairs. Both groups seem to require a similar amount of time with both an IDE and a simple text editor. With respect to the time required to code both programs, the calculator program required three-times as much time than did the encoder program.

The descriptive statistics provide an overview of the collected measurements; however, at this point, we are unable to draw any conclusions with confidence with regard to possible differences between treatments. Once we have an overview of the data, we will proceed with the inferential statistics to test the previously stated hypotheses. The statistical model employed according to the factorial design with repeated measurements is defined in Equation (1).

$$y_{ijkl} = \mu + \alpha_i + \beta_j + d_{ijl} + \gamma_k + (\alpha\beta)_{ij} + (\alpha\gamma)_{ik} + (\beta\gamma)_{jk} + (\alpha\beta\gamma)_{ijk} + \varepsilon_{ijkl} \ . \qquad (1)$$

where $\mu$ is the general mean, $\alpha_i$ is the effect of the $i$th treatment (programming type), $\beta_j$ is the effect of the $j$th treatment (tool support), $d_{ijl}$ is the random experimental error for the experimental units (pairs and solos) within treatments (programming type and tool support) with variance $\sigma^2_d$, $\gamma_k$ is the effect of the kth program, $(\alpha\beta)_{ij}$ is the interaction between the programming type and the program, $(\alpha\gamma)_{ik}$ is the interaction between the programming type and the program, $(\beta\gamma)_{jk}$ is the interaction between the tool support and the program, and $(\alpha\beta\gamma)_{ijk}$ is the interaction between the programming type, tool support and program. Finally, $\varepsilon_{ijkl}$ is the normally distributed random experimental error on repeated measurements with variance $\sigma^2_e$.

To assess each of the components of this statistical model, an analysis of variance (ANOVA) is applied. ANOVA relies on an analysis of the total variability of the collected measurements and the variability partition according to different components (in this case, factors and their interactions). ANOVA provides a statistical test of whether the means of several groups (of measurements) are all equal. The null hypothesis is that all groups are simply random samples of the same population. This implies that all treatments have the same effect (perhaps none). Rejecting the null hypothesis implies that different treatments result in different effects. In this experiment, we have four groups of measurements (programming type and tool support combinations), with two repeated measurements per group (the two programming assignments).

ANOVA was applied through the use of the R package ez [41], which implements a function for the analysis of factorial designs with repeated measurements. Table VI shows the ANOVA results with respect to product quality.

TABLE VI
ANOVA results for product quality

| Component | DFn | DFd | F | $p$ | $p<0.05$ |
|---|---|---|---|---|---|
| Programming type | 1 | 12 | 0.939677 | 0.351493 | |
| Tool support | 1 | 12 | 7.913204 | 0.015660 | * |
| Program | 1 | 12 | 13.491128 | 0.003189 | ** |
| Prog. type:Tool support | 1 | 12 | 2.308808 | 0.154542 | |
| Prog. type:Program | 1 | 12 | 1.336983 | 0.270068 | |
| Tool support:Program | 1 | 12 | 4.264636 | 0.061216 | . |
| Prog. type:Tool:Program | 1 | 12 | 0.152236 | 0.703244 | |

As shown in Table VI, the tool support and program components exhibit a significant difference at an alpha level of 0.05 and 0.01, respectively, suggesting that defect injection rates are different for different types of tool support (IDE and simple text editor) and for different programming assignments (encoder and calculator). The interaction between tool support and program shows certain level of significance (it is close to an alpha level of 0.05). However, if we set an alpha level of 0.1, which represents a confidence level of 90%, this interaction exhibits a significant difference.

Estimating the effect sizes for these components, a generalized $\eta^2$ [42] of 0.38, 0.23 and 0.08 was observed for tool support, program, and the interaction between tool support and program, respectively. According to [43], these effect sizes can be interpreted as large ($\eta^2>0.14$), large and medium ($\eta^2>0.06$), respectively.

Because a significant interaction was observed, we conduct a post-hoc test with multiple pairwise comparisons with the goal of examining differences in all the level combinations between tool support and program. Table VII shows the post-hoc test results using Tukey's method [44].

TABLE VII
Pairwise comparisons between tool support and program

| Comparison | Estimate | Std. Error | z value | Pr(>\|z\|) | |
|---|---|---|---|---|---|
| Text editor.Calculator − IDE.Calculator =0 | -1.3362 | 1.1550 | -1.157 | 0.64122 | |
| IDE.Encoder − IDE.Calculator =0 | 3.4642 | 0.8457 | 4.096 | < 0.001 | ** |
| Text editor.Encoder − IDE.Calculator =0 | -0.4509 | 1.1550 | -0.390 | 0.97862 | |
| IDE.Encoder − Text editor.Calculator =0 | 4.8004 | 1.1550 | 4.156 | < 0.001 | ** |
| Text editor.Encoder − Text editor.Calculator =0 | 0.8853 | 0.7459 | 1.187 | 0.62198 | |
| Text editor.Encoder − IDE.Encoder =0 | -3.9151 | 1.1550 | -3.390 | 0.00356 | ** |

According to the results in Table VII, three pairwise comparisons exhibit significant differences (at α=0.01). In the first one, a significant difference of 3.5 defects injected per hour is observed between the encoder and calculator programs, both programs coded with an IDE. In the second one, a significant difference of 4.8 defects injected per hour is observed between the encoder program coded with an IDE and the calculator program coded with a text editor. In the third one, a decrease of 3.9 defects injected per hour is observed when a text editor is used to code the encoder program. A visual representation of this interaction (tool support and program) is given in Fig. 1.

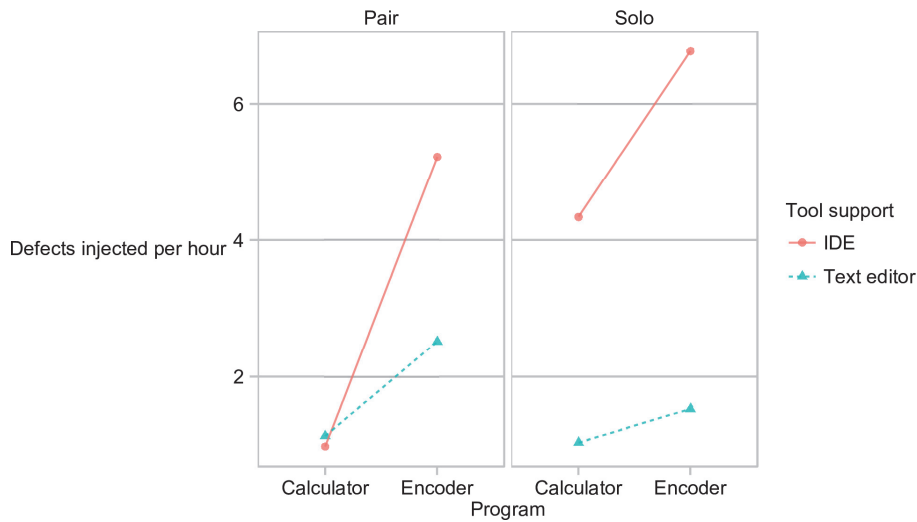

Fig. 1. Interaction plot between tool support and program with regard to quality

As shown in Fig. 1, the use of an IDE for the encoder program produces higher defect injection rates than does the use of a text editor. In the case of the calculator program, the use of an IDE appears to reduce the defect injection rates, but only for pairs. Another representation of the defect injection rates between programming type and tool support categorized by programming assignment is shown in Fig. 2.

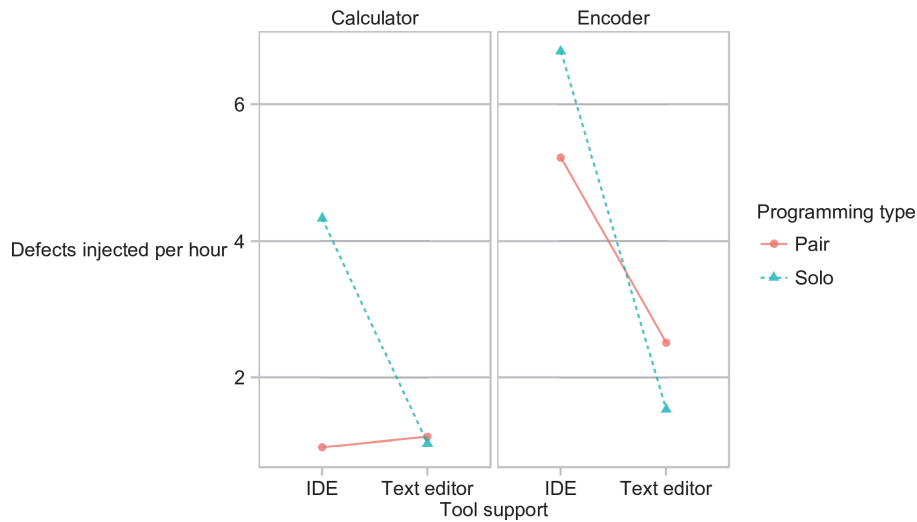

Fig. 2. Interaction plot between programming type and tool support with respect to quality

According to this figure (Fig. 2), solo programmers exhibit a similar pattern when coding both programs, and the use of a text editor seems to significantly reduce the defect injection rates in comparison to the use of an IDE. This is partially supported for pairs working on the encoder program. This appears to be an interaction between programming type and tool support (although not significant), wherein pairs seem to inject fewer defects when using an IDE compared to solos. Conversely, solos seem to inject fewer defects when using a simple text editor compared to pairs. Regarding the program, a small number of defects were injected into the calculator program. In this program, pairs produce similar defect injection rates with either the use of an IDE or a text editor.

With regard to cost, measured as the time (in minutes) that participants spend coding the programming assignments, the analysis of variance is shown in Table VIII. For this aspect, the program component exhibits a significant difference (at an alpha level of 0.01), therein suggesting a significant difference between the coded programs. This component shows an effect size $\eta^2$ of 0.28, which is interpreted as a large effect size [43].

TABLE VIII
ANOVA results for cost

| Component | DFn | DFd | F | p | p<0.05 |
|---|---|---|---|---|---|
| Programming type | 1 | 12 | 0.293834 | 0.597695 | |
| Tool support | 1 | 12 | 0.007055 | 0.934445 | |
| Program | 1 | 12 | 15.348790 | 0.002043 | ** |
| Prog. type:Tool support | 1 | 12 | 0.000292 | 0.986640 | |
| Prog. type:Program | 1 | 12 | 0.034953 | 0.854816 | |
| Tool support:Program | 1 | 12 | 0.346383 | 0.567080 | |
| Prog. type:Tool:Program | 1 | 12 | 0.000002 | 0.998689 | |

This significant difference is shown in Fig. 3; participants spent less time coding the encoder program than they did coding the calculator program. Pairs tend to require less time than do solos, although this difference is not significant. The use of an IDE seems to reduce the coding time for the encoder program; however, it increases when an IDE is used for coding the calculator program. This interaction is not significant, though.

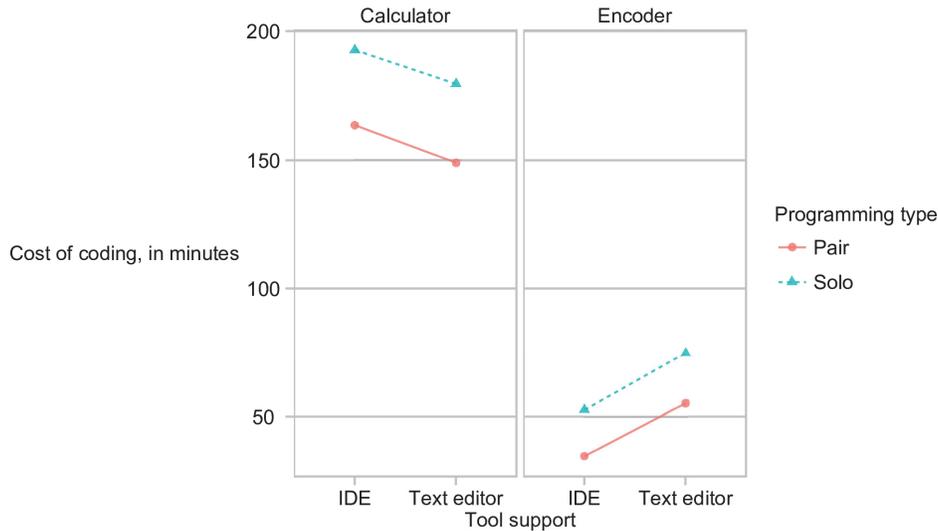

Fig. 3. Interaction plot between programming type and tool support with respect to cost

To draw valid statistical conclusions, the employed statistical model was checked against the assumptions of normality, sphericity and randomness. One way of assessing normality is by examining whether the collected measurements conform to a normal distribution at each level of the within-subject factor. For the standardized residuals of each level of the within-subject factor, we used the Jarque–Bera test for normality [45,46]. The null hypothesis of this test assumes that sample data originate from a normal distribution. Regarding quality, the standardized residuals for the encoder program show a p-value of 0.3202, and for the calculator program, a p-value of 0.0004 is found. Regarding cost, the standardized residuals for the encoder program show a p-value of 0.1133, and for the calculator program, a p-value of 0.1823 is found.

With regard to quality, the assumption of normality is violated for the calculator program. The observed outlier belongs to $eu_{14}$, which has the smallest defect injection rate in the programming type and tool support treatment combination, although this experimental unit registered a higher than average number of defects (the average for this program was 3.18). This unit spent 479 minutes coding this program. On the other hand, for the encoder program, the null hypothesis is accepted in favor of normality. Regarding cost, the null hypothesis is accepted in favor of normality for both programs.

The sphericity assumption implies that the variances of the differences between any two levels of within-subject factors (factors repeated over the same experimental unit) are similar. This assumption is always met when there are only two levels of within-subject factors [47], i.e., two repeated measurements, as in our case, where the two repeated measurements were for the encoder and calculator programs. With respect to the randomness assumption, the collected measurements were sampled randomly and independently of each other.

It is important to note that the previous assumptions are related to a univariate statistical model similar to the one described in Equation (1). However, a repeated measurement model can be seen as a multivariate statistical model wherein the repeated measurements act as dependent (or response) variables. Under a multivariate statistical model, two common assumptions to assess are the multivariate normality and the homogeneity of covariance matrices [48].

*A.      Qualitative results*

After the experiment sessions were completed, participants working in pairs responded to a questionnaire regarding their perceptions of the use of pair programming. The questionnaire was responded to individually.

The first question referred to the degree of cohesion achieved (team jellying) during all experimental sessions. We defined a scale from 1 to 9, where 1 represents the lowest level of cohesion achieved and 9 the highest level of cohesion achieved. On average, the participants perceived a good level of cohesion with their respective partners (8.18, SD=0.73).

For the second question, we asked whether the cohesion was increased or decreased during the experimental sessions. With the exception of one respondent (6%), who answered to have the same level of cohesion during the sessions, the rest of respondents (94%) stated that the perceived cohesion increased during the sessions.

For the third question, we asked whether the participant would work on future assignments with the same partner given the achieved cohesion. 70% of the respondents would work on future assignments with the same partner, 12% would not and 18% perhaps would do it.

The fourth question was related to the role that participants were mainly involved in. 35% of the respondents stated the controller as being their main role, 41% acted as monitors and 24% responded as being equally involved in both roles.

Concerning whether participants changed role during the experiment (fifth question), 59% of the respondents answered that they did change roles, whereas 41% maintained the same role. Finally (sixth question), all respondents stated to have enjoyed the programming role that they were involved in.

## V.   Discussion

Now that the analysis and results have been presented, in this section, we discuss them in relation to the hypotheses defined and with previous work.

In terms of quality, which is measured as the number of defects injected per hour, pairs and solos seem to develop software with a similar level of quality (**H0$_a$** is accepted). Coding using an IDE and a simple text editor yield software products with different levels of quality (**H0$_b$** is rejected); the use of an IDE seems to generate higher defect injection rates than does the use of a simple text editor. The programming assignments being coded yield software products with different levels of quality (**H0$_c$** is rejected); in this case, participants coding the program identified as encoder produced higher defect injection rates compared to those coding the calculator program. Software product quality is not affected by the relationship between the type of programming (pair or solo) and the type of tool support used (**H0$_d$** is accepted). Software product quality is not affected by the relationship between the type of programming and the type of program being coded (**H0$_e$** is accepted). Software product quality is affected by the relationship between the type of tool support used and the type of program being coded (**H0$_f$** is rejected). Software product quality is not affected by the relationship between the type of programming, tool support and programming assignment (**H0$_g$** is accepted).

Because a significant interaction was observed (at $\alpha$=0.1) between tool support and the programming assignment, interpretations should be based on interaction effects and not on main effects. A significant difference is observed for the programming assignment when it is coded with an IDE; the use of an IDE produced greater defect injection rates for the encoder program. Another significant difference is observed for the calculator program coded with a simple text editor and the encoder program coded with an IDE; the latter combination (IDE-encoder) yielded the highest defect injection rates. A third significant difference is observed for tool support and the encoder program. In this programming assignment, the use of a simple text editor produced lower defect detection rates.

Although on average the calculator program presents a lower cyclomatic complexity (VG =6.93, SD=7.79) compared to the encoder program (VG =16.25, SD=19.01), the calculator program is more complicated to code in the sense that it demands greater effort to be implemented. This programming assignment implies certain knowledge on how to address regular expressions.

According to our results, it seems that pairs, irrespective of the tool support that they use, tend to produce software products with better quality when they work on complex coding tasks (such as the calculator program used in our experiment); in this case, the utilized tool support seems to have no effect (as shown in Fig. 1). Our findings reinforce those reported in [49, 16], wherein has been observed than junior pairs perform better when they work on more complex programs [49], and also has been observed that pair programming performs well when novice pairs encounters challenging programming problems. On the other hand, when coding tasks are less complicated, (such as the encoder program used in our experiment), pairs and solos seem to produce software with better quality when a simple text editor is used.

In terms of cost, which is measured as the time required in minutes to complete the programming assignment, pairs and solos seem to require similar amounts of time to code the programming assignments (**H0$_h$** is accepted). The cost of coding using an IDE and that using a simple text editor are equivalent (**H0$_i$** is accepted). The programming assignments require different amounts of time to be coded (**H0$_j$** is rejected); in this case, participants coding the encoder program required less time. Cost is not affected by the relationship between the type of programming (pair or solo) and the type of tool support used (**H0$_k$** is accepted). Cost is not affected by the relationship between the type of programming and the type of program being coded (**H0$_l$** is accepted). Cost is not affected by the relationship between the type of tool support used and the type of program being coded (**H0$_m$** is accepted). Cost is not affected by the relationship between the type of programming, tool support and the programming assignment (**H0$_n$** is accepted).

With respect to related work, our findings are compared only with regard to cost. This is because effort is not analyzed in the present work and because product quality is not analyzed in [10]. Regarding cost, our results suggest that pairs complete programming assignments in less time than do solos, but this difference is not significant. Conversely, the authors in [10] observed a significant difference (at $\alpha$=0.1) in favor of pairs, whom also completed the assignment in less time than did solos. In this sense, our results are similar to those reported in [49-51,17,52-55], wherein the authors did not observe significant differences when applying the pair programming practice.

In reference to the programming assignments, our findings reinforce those reported in [10]; the calculator program requires more time to code than does the encoder program. Regarding tool support, our findings are also congruous with those in [10]; the cost of coding is similar when using an IDE compared to when using a text editor.

Because tool support and program where used as blocking factors in [10], it was no possible to assess possible interaction effects; however, this was alleviated with the experimental design used in this work. This design allowed the assessment of interactions between treatments of interest regarding quality and cost.

*A.    Study Limitations*

Empirical studies are subject to undermining threats. Next, we describe strategies that we followed to minimize the threats to validity [56]. With respect to conclusion validity, the measurements collected during the experiment satisfy the principles of normality[1], sphericity and randomness. With respect to internal validity, the participants were randomly assigned to treatments, which reduced learning effects. Boredom or fatigue was reduced by using alternating training and experimental sessions. The experimental units were placed in the same classroom, worked under the same conditions, and sat apart, with no interaction. With respect to construct validity, cause constructs were operationalized in the same manner as in previous studies on the topic; effect constructs were operationalized in the same manner as in [57] (for quality) and as in [10] (for cost). With respect to external validity, the use of students instead of practitioners might have compromised this type of validity. However, there is evidence that suggests that in some contexts, the results of empirical studies that employ students with sufficient technical skills can be equivalent to the results of empirical studies that use practitioners [58,59]. In this respect, the participants in this experiment reported having almost two years of experience with the Java programming language and almost two years of experience with the NetBeans IDE. In the context of pair programming studies, there is evidence in favor of this claim: pair programming studies that employ practitioners [49,54] report findings that are similar to studies employing students [50, 55].

## VI.    CONCLUSIONS

In this work, we presented a controlled experiment that assessed the quality and cost of pair and solo programming with and without the use of an integrated development environment (IDE). Although different aspects of pair programming have been studied, there is no work that examines the effects of using or not using a tool support such as an IDE under a traditional model of pair programming, in which pairs work at the same location at the same time with the same computer.

In terms of quality, our results suggest that the use of an IDE produces higher defect injection rates (for both pairs and solos) when the programming assignment is not very complicated. Nevertheless, defect injection rates seem to decrease when pairs work on more complicated programming assignments, irrespective of the tool support that they use.

At first glance, instinct may suggest that the use of an IDE is more effective than the use of a text editor; however, the evidence presented here suggests the opposite. One possible reason for our findings is that when entry-level programmers use an IDE, it is common for programmers to write some lines of code and then compile and run it. It is possible that this mechanism interrupts the concentration flow required for coding. On the other hand, when a text editor is used for coding, programmers tend to prolong the compiling and running process; they have to use a console and manually perform these operations. It is possible that this mechanism encourages a longer concentration flow, which in our case produced lower defect injection rates. In an academic context, these findings suggest that students enrolled in programming courses should have the use of a text editor reinforced for a certain period of time with the goal of obtaining greater concentration while coding.

In terms of cost, the programming assignment effects the time required to complete the assignment. Finally, regarding both studied aspects (quality and cost), pairs and solos behave in a similar manner. Future replications of this experiment need to be conducted to gain additional insight into the findings presented here.

---

[1] Particularly, in the case of cost, normality is satisfied; for quality, normality is assumed only for the encoder program. In reference to the calculator program, a departure from normality was observed due to the presence of an outlier, as discussed in Section 4.